\documentclass[a4paper,10pt,twoside,superscriptaddress]{cpc-hepnp}
\usepackage{amssymb}
\usepackage{CJK,upgreek,fancyhdr}
\usepackage{multicol}
\usepackage{graphicx}
\usepackage{subfigure}
\usepackage{booktabs}
\usepackage{amssymb,bm,mathrsfs,bbm,amscd}
\usepackage[tbtags]{amsmath}
\usepackage{lastpage}
\usepackage{epsfig}
\usepackage[colorlinks,linkcolor=blue,anchorcolor=blue,citecolor=blue,dvipdfm]{hyperref}
\usepackage{amsmath}
\usepackage{mathrsfs}
\usepackage{dcolumn}
\usepackage{footnote}
\usepackage{booktabs}
\usepackage{threeparttable}
\usepackage{multirow}

\begin{document}
%\begin{CJK*}{GB}{gbsn}
\begin{CJK*}{GBK}{song}
%\linenumbers
%\fancyhead[c]{\small Chinese Physics C~~~Vol. xx, No. x (2016) xxxxxx}
%\fancyfoot[C]{\small 010201-\thepage}

%\footnotetext[0]{Received xx June 201x}

\title{Intrinsic radiation background of LaBr$_3$(Ce) detector via coincidence measurements and simulations\thanks{Supported by National Natural Science
Foundation of China (11575018, 11375023, 11305269, 11375267, 11405274, 11205245, 10927507, 10975191, 11075214, 11175259, 11205068)}}

\author{%
      Hao Cheng $^{1,2,3)}$
\quad Bao-Hua Sun $^{1,2,3)}$\email{bhsun@buaa.edu.cn}%
\quad Li-Hua Zhu $^{1,2,3)}$\email{zhulh@buaa.edu.cn}%
\quad Tian-xiao Li $^{3)}$\\
\quad Guang-Shuai Li $^{1)}$
\quad Cong-Bo Li $^{3)}$
\quad Xiao-Guang Wu $^{3)}$
\quad Yun Zheng $^{3)}$
}
\maketitle

\address{%
$^1$ School of Physics, Beihang University, Beijing 100191, China\\
$^2$ Beijing Advanced Innovation Center for Big Data-Based Precision Medicine, \\School of Medicine and Engineering, Beihang University, Beijing, 100191, China; \\and Key Laboratory of Big Data-Based Precision Medicine (Beihang University), Ministry of Industry and Information Technology\\
$^3$ China Institute of Atomic Energy, Beijing 102413, China\\
}

\begin{abstract}
The LaBr$_3$(Ce) detector has attracted much attention in recent years for its
superior characteristics to other scintillating materials in terms of resolution and efficiency.
However, it has relatively high intrinsic radiation background due to the naturally occurring radioisotope in lanthanum, actinium and their daughter nuclei. This limits its applications in low counting rate experiments.
In this paper, we identified the radioactive isotopes in the $\phi3''\times 3''$ Saint-Gobain B380 detector by a coincidence measurement using a Clover detector in a low-background shielding system.
Moreover, we carried out a Geant4 simulation to the experimental spectra to evaluate the activities of the main internal radiation components. % by means of the $\alpha$, $\gamma$ and $\beta$ decays from the $^{138}$La and $^{227}$Ac.
The activity of radiation background of B380 is determined to be 1.480 (69) Bq/cm$^3$, the main sources of which include $^{138}$La of 1.425 (59) Bq/cm$^3$, $^{211}$Bi of 0.0136 (15) Bq/cm$^3$, $^{219}$Rn of 0.0125 (17) Bq/cm$^3$, $^{223}$Ra of 0.0127 (14) Bq/cm$^3$, and $^{227}$Th of 0.0158 (22) Bq/cm$^3$.
\end{abstract}


%\begin{pacs}
%21.10.Re, 27.60.+j, 21.60.Ev
%\end{pacs}

%\footnotetext[0]{\hspace*{-3mm}\raisebox{0.3ex}{$\scriptstyle\copyright$}2013
%Chinese Physical Society and the Institute of High Energy Physics of the Chinese Academy of Sciences and the Institute of Modern Physics of the Chinese Academy of Sciences and IOP Publishing Ltd}%

\begin{multicols}{2}

\section{Introduction}\label{intro}

As a new type of inorganic scintillation, the LaBr$_3$(Ce) crystal has a high density of 5.08 g/cm$^3$, a high light output of about 63 photons/keV$\gamma$, a fast decay time of about 16 ns~\cite{saintgobain} and a good temperature response.
These superior characteristics make LaBr$_3$(Ce) ideal for many applications~\cite{Iltis2006Lanthanum,Liu2012Using,Young2018Application,Mouhti2018Validation,Cheng2018Determination} in environmental monitoring, oil well logging, nuclear safeguards, and medical imaging. LaBr$_3$(Ce) thus is often taken as a substitution to the widely used NaI(TI) crystal when high performance is demanded. The integrated LaBr$_3$(Ce) detector consists of a crystal coupled directly to a selected photomultiplier tube (PMT). Previous studies of LaBr$_3$(Ce) detectors~\cite{saintgobain,Shah2003LaBr3,Nicolini2007Investigation,munal,Quarati2007X,Van2000High,Lavagno2013Study,M2013Improvement,Shi2018Unfolding} show excellent linearity in $\gamma$ ray response, a good energy resolution of less than 3\% (FWHM) for the 662 keV $\gamma$ ray for the size of up to  $\phi3''\times3''$, and an excellent time resolution of about $\sim$300 ps (FHWM). The latter has make a fast timing detector array composed of LaBr$_3$(Ce)~\cite{Alison2014A,Longfellow2019Commissioning} pursued worldwide for nuclear structure studies.

On the other hand,  LaBr$_3$(Ce) detector has a relatively high intrinsic radiation background~\cite{Milbrath2005Characterization,Milbrath2006Contamination,Rosson2011RADIATION,Quarati2012Study,Lavagno2013Study,Camp2016Determination}, which is typically more than 1 to 2 orders of magnitude higher than that of NaI(Tl) detector.
The self-radiations root in $^{138}$La and the five short-lived progeny of $^{227}$Ac impurities, may cause a non-negligible effect in the energy spectrum as a result.
This would seriously limit its application in low count rate experiments such as space $\gamma$ rays.
Therefore, it is valuable to quantify the intrinsic radiation of LaBr$_3$(Ce) and moreover to understand their influence.

The present study aims to identify the components of internal radiation in LaBr$_3$(Ce)  and furthermore to deduce their activities. The detector of interest is Saint-Gobain B380 with the size of $\phi3''\times3''$.
This is done by combining the coincidence measurement with the dedicated Geant4 simulation.
The paper is organized as follows. Section \ref{exp}  presents the coincidence measurement of LaBr$_3$(Ce) vs. a Clover detector, and the corresponding results. In Sec. \ref{dis}
we made Geant4 simulations to the experimental spectra of both LaBr$_3$(Ce) and Clover detector, and deduce the activity of $^{138}$La , $^{221}$Bi, $^{219}$Rn, $^{223}$Ra and $^{227}$Th.
A summary is given in Sec.  \ref{sum}.

\section{Internal radiation of LaBr$_3$(Ce) and coincidence measurement}\label{exp}

\subsection{Internal radiation of LaBr$_3$(Ce) detector}

$^{138}$La is the only naturally occurring radioactive isotope of lanthanum with 0.09\% abundance and a half-life of $1.05\times 10^{11}$ years, which affects the energy spectrum below 1.5 MeV.
$^{138}$La decays in two parallel processes, as shown in Fig. ~\ref{fig1}.  34.4\% of the isotope undergoes $\beta^-$ decay, with a maximum energy of 263 keV, eventually to the first excitation state of $^{138}$Ce. The process is associated with the emission of a 788.742 keV $\gamma$ ray. The remaining 65.6\% of $^{138}$La disintegrates by electronic capture (EC). This process results in stable $^{138}$Ba with the emission of a 1435.795 keV $\gamma$ ray and the characteristic X-rays of Ba with energies ranging from 31 to 38 keV.
%%%%%%%%%%%%%%%%%%%%%%%%%%%%%%%%%%
\begin{center}
\includegraphics[width=8.0cm,clip=true,trim=0cm 0cm 0cm 0cm]{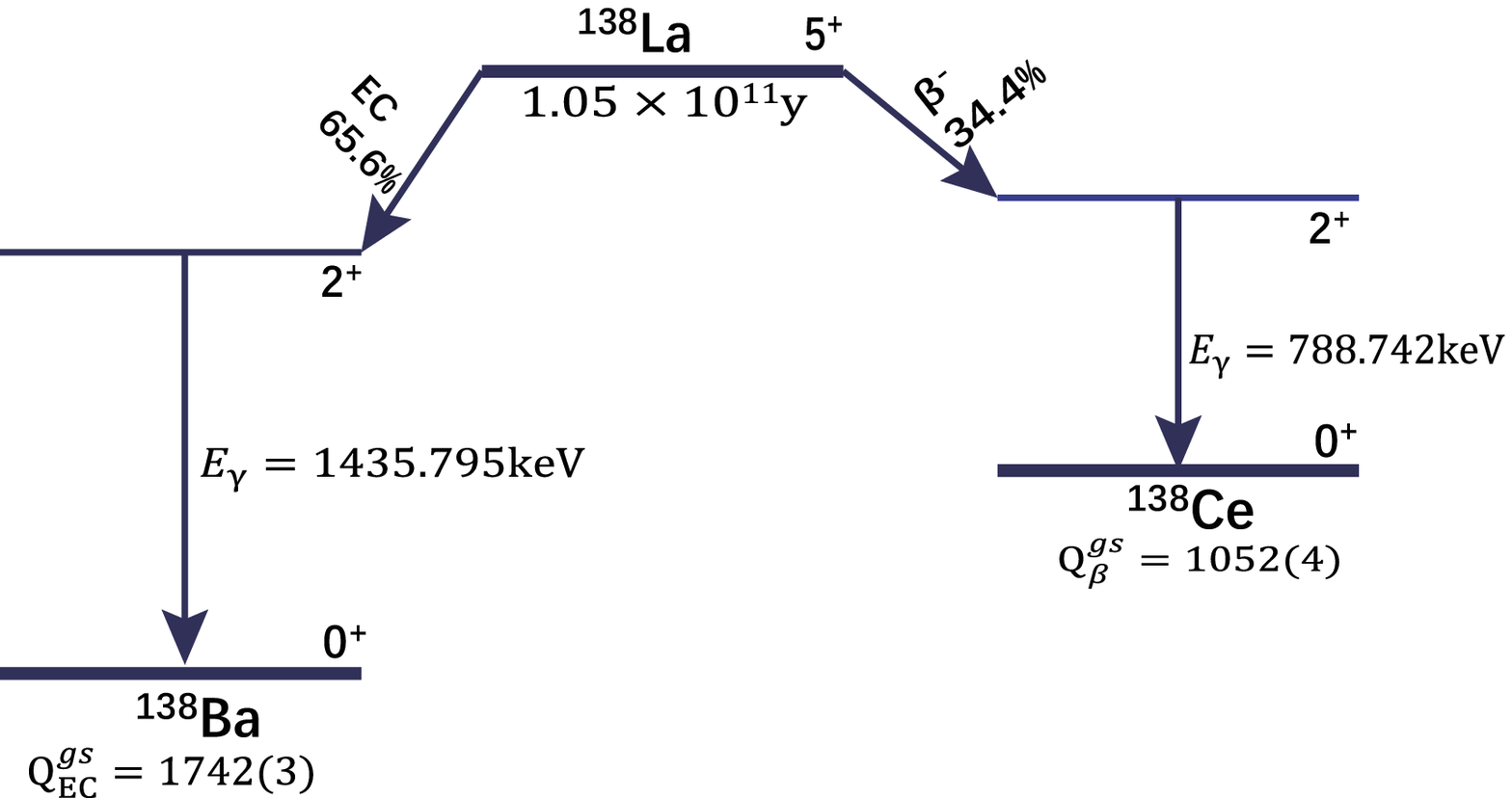}
\figcaption{\label{fig1} The decay scheme of  $^{138}$La. The data are from NNDC~\cite{NNDC}.}
\end{center}
%%%%%%%%%%%%%%%%%%%%%%%%%%%%%%%%%%%

$^{227}$Ac is the grand-daughter nuclide in the $^{235}$U decay chain.
Due to the similarity in chemistry to lanthanum, it presents as a contaminator with a half-life of of 21.77 years
in LaBr$_3$(Ce).  Fig.~\ref{fig2} shows the decay chain down to the stable $^{207}$Pb by emitting $\alpha$, $\beta$ and $\gamma$ rays. Here includes 6 $\alpha$ emitters ($^{227}$Ac, $^{227}$Th, $^{223}$Ra, $^{219}$Rn, $^{215}$Po and $^{211}$Bi), and 4 $\beta$ emitters ($^{227}$Ac, $^{211}$Pb, $^{211}$Bi and $^{207}$Tl).  $^{227}$Ac and its daughter nuclei produce the higher energy background by emitting $\alpha$ particles, and also contribute to the $\beta$ continuum up to about 1400 keV due to the $\beta$ decay of $^{211}$Pb and $^{207}$Tl in the decay chain of this nucleus.
%%%%%%%%%%%%%%%%%%%%%%%%%%%%%%%%%%%
%%%%%%%%%%%%%%%%%%%%%%%%%%%%%%%%%%
\begin{center}
\includegraphics[width=8.0cm,clip=true,trim=0cm 0cm 0cm 0cm]{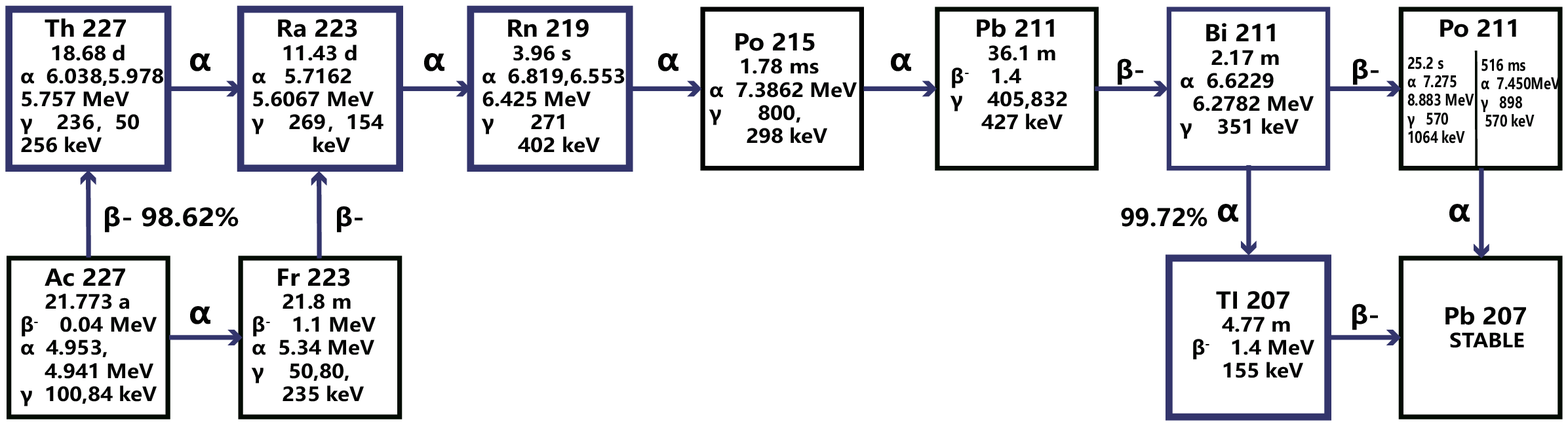}
\figcaption{\label{fig2} Actinide decay chain. Listed are the half-life,  energies of $\alpha$ and characteristic $\gamma$ rays with relatively high intensities and $\beta$-decay end-point energy for each nuclide. Data are taken from NNDC~\cite{NNDC}. }
\end{center}

%%%%%%%%%%%%%%%%%%%%%%%%%%%%%%%%%%%
\subsection{Experimental setup}

The coincidence measurement using a HPGe was conducted in a low-background counting system (LBS).
The environmental background counting rate is 58 per second. The LBS is a cylinder with a radius of 64 cm and a height of 66.1 cm, and consists of four layers: iron, lead, copper and plexiglass from the outside to the inside. The schematic diagram of the entire detection system is shown in Fig. ~\ref{fig3}. The lead layer can shield most of the low energy environment background, and the copper layer aims to absorb the characteristic X-rays of lead.
\begin{center}
\includegraphics[width=5.0cm,clip=true,trim=0cm 0cm 0cm 0cm]{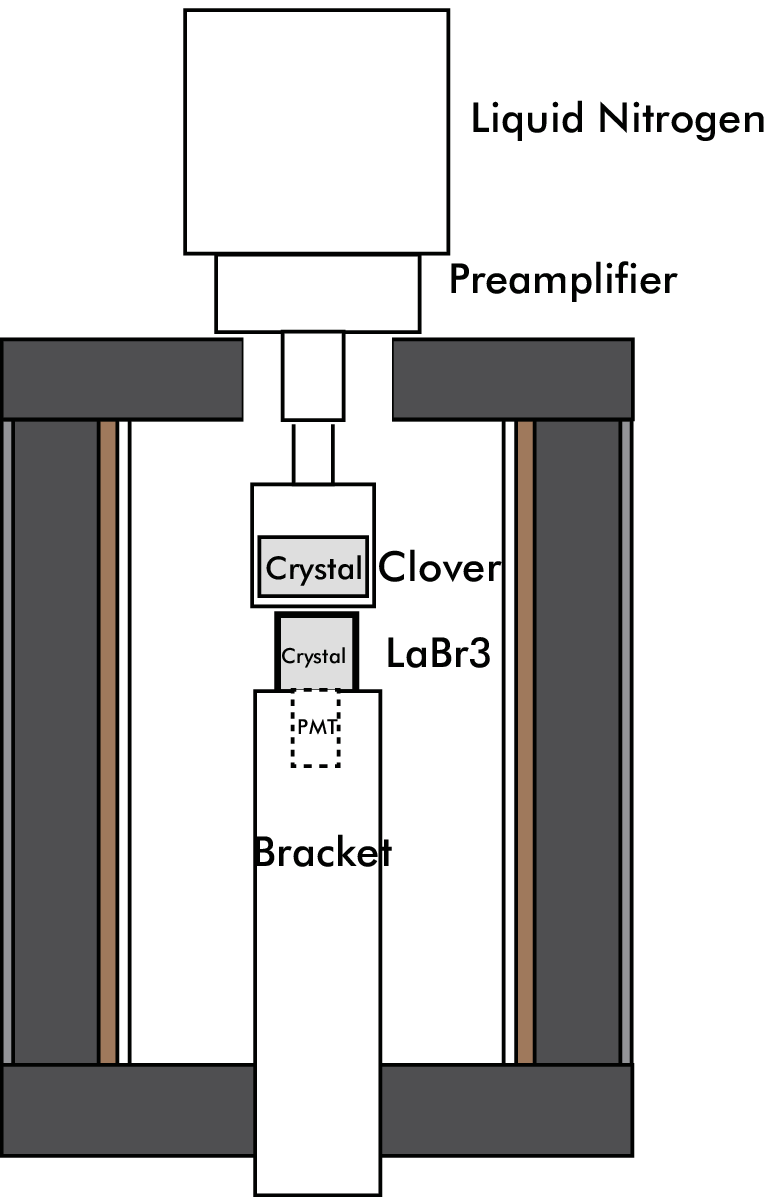}
\figcaption{\label{fig3}(Color online) Schematic diagram of experimental setup. It contains LaBr$_3$ and Clover crystals, and a low-background shielding room. The shielding room is composed of plexiglass, copper, lead and iron from the inside to the outside. The LaBr$_3$ detector is supported by a bracket in the shielded room. }
\end{center}

The LaBr$_3$(Ce) detector is the Saint-Gobain B380 with a $\phi3^{''}\times3^{''} $ crystal.
A high-purity germanium (HPGe) Clover detector of Canberra was placed directly facing to the LaBr$_3$(Ce).
The high voltage applied for the LaBr$_3$(Ce) detector was set to be 520 V. Higher voltage may cause the electron saturation in the photomultiplier, thus can affect linearity~\cite{Trofimov2013Linearity} in the energy determination.
The Clover consists of four coaxial N-type high-purity Germanium detectors, each with a diameter of 60 mm and length of 60 mm.
The energy resolutions for the  LaBr$_3$(Ce)  and Clover are measured to be 2.1\% (FWHM) and 0.166\% (FHWM) for the 1.332 MeV $\gamma$ ray, respectively.  The relative efficiency is 38\% for each crystal.

The energy and time signals of the two detectors were acquired by the VME data acquisition system collected for 37,187 seconds in total. Dead time correction and time stamp were added in the data acquisition software. Standard radioactive sources $^{60}$Co, $^{137}$Cs, and $^{241}$Am were used for the $\gamma$-ray energy calibration up to around 1.33 MeV. The calibration accuracy
has been cross-checked by the characteristic $\gamma$-rays of $^{138}$La. Moreover, to calibrate the high energy spectrum of LaBr$_3$(Ce),
we have used the recoil-electron energies from the Compton scattering process of 2.615 MeV $\gamma$ ray of
$^{208}$Tl, a naturally radioactive nuclide in environmental background. Such calibration is only possible by using the coincidence measurement with the Clover.

\subsection{Coincidence measurement}

%%%%%%%%%%%%%%%%%%%%%%%%%%%%%%%%%%%
\begin{center}
\includegraphics[width=9.0cm,clip=true,trim=0cm 0cm 0cm 0cm]{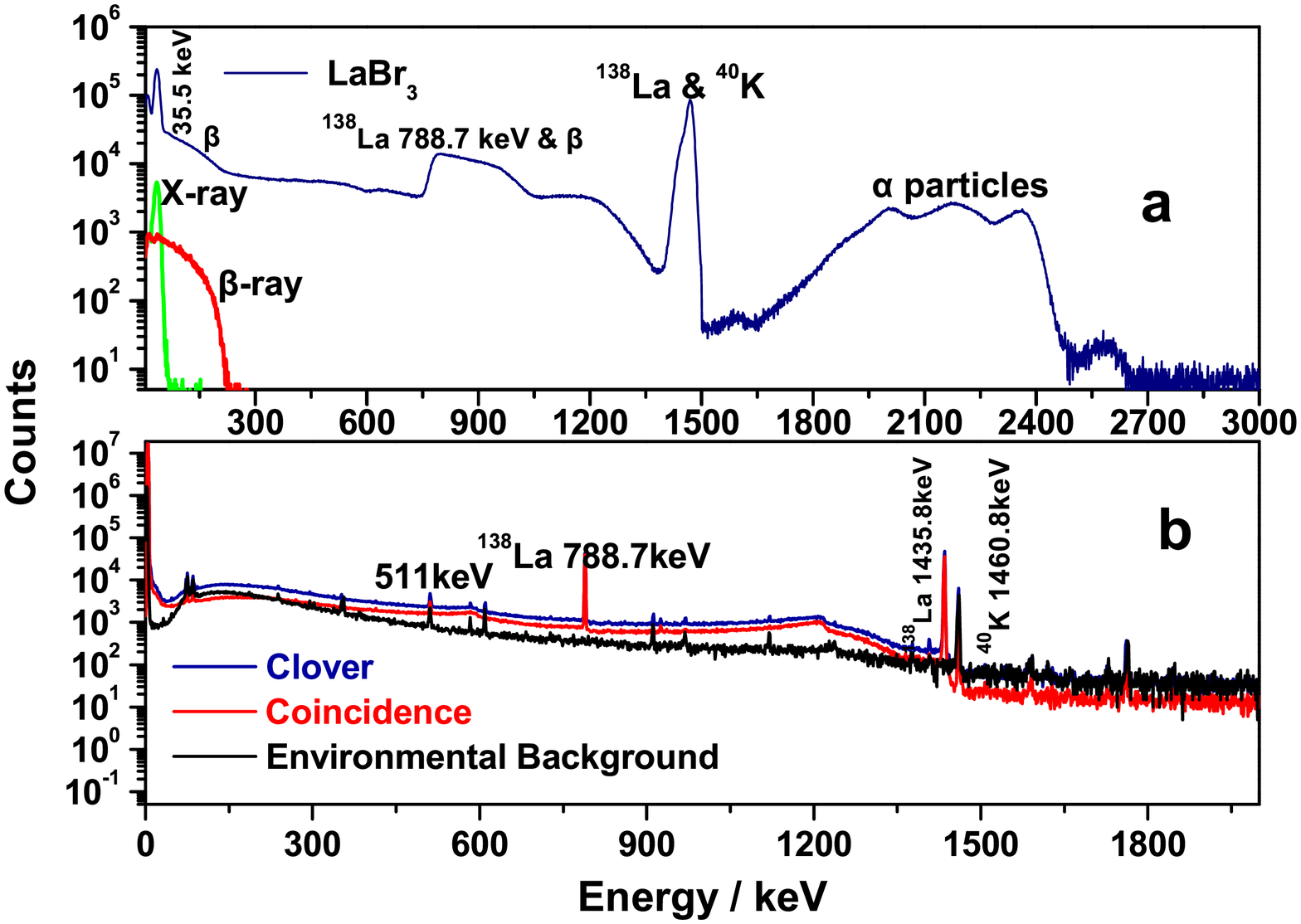}
\figcaption{\label{fig4}(Color online) Radiation background spectra measured in 37,187 seconds by the LaBr$_3$(Ce) detector (a) and the Clover (b). The coincidence $\beta$ spectrum (red) and X-ray spectrum (green) in the LaBr$_3$(Ce) detector are shown in (a). Coincidence spectrum (red) and environment background spectrum (black) of Clover  are also shown in (b). For details, refer to the text.}
\end{center}

%%%%%%%%%%%%%%%%%%%%%%%%%%%%%%%%%%
 The intrinsic radiations of the LaBr$_3$(Ce) scintillator were identified by the coincidence measurement of the LaBr$_3$(Ce) and Clover detectors.
The relevant background spectra after energy calibration are shown in Fig.~4.
%The top part of  Fig. 4 shows the self-irradiation energy spectrum of the LaBr$_3$(Ce) detector.
In the self-counting spectrum of LaBr$_3$(Ce), we see first a low-energy peak centered at around 35.5 keV. It is attributed to the sum of  95.6$\%$ of the 31.83 keV K$_\alpha$ X-ray response and 90 $\%$ of the 5.6 keV Auger electron response in the EC decay process, by referring to the theoretical calculation~\cite{Quarati2012Study}. The energy shift is due to the non-proportional response of the LaBr$_3$(Ce) crystal.
Then we see a $\beta$ continuum with an end point of 263 keV mixed with the Compton continua from mainly 788.7, 1435.8 keV of $^{138}$La and the 1460.8 keV of $^{40}$K.
With the increasing energy, the 788.7 keV bump is shown to extend to higher energies and end at around 1 MeV.
It is due to the coincidence of 788.7 keV $\gamma$ with the $\beta^-$ continuum.
The 1435.8 keV $\gamma$-rays produced by the EC of $^{138}$La coincided with the 32 keV X-rays of $^{138}$Ba and the 1460.8 keV $\gamma$-rays of $^{40}$K, result in a double peak near 1461 keV, as shown in Fig.~\ref{fig4}(a).

The spectrum above 1.5 MeV shows a three-peak structure, revealing the presence of $\alpha$ emitter contaminants.
Although the $\alpha$ energies from $^{227}$Ac and its daughter nuclei are as high as 5.0$\sim$7.4 MeV (see Fig.~\ref{fig2}),
the energies are read out in the spectrum calibrated with $\gamma$-rays to be in the energy range of 1.5 and 2.5 MeV due to the well-known light quenching effect (see, e.g. Ref.~\cite{Milbrath2006Contamination}).
The species will be further explored later by the coincidence measurement. % using a HPGe.

The energy spectrum of the Clover detector as well as the coincidence spectrum are shown in the Fig.~\ref{fig4}(b). It is clearly seen that the characteristic $\gamma$-rays of 788.7 keV and 1435.8 keV of $^{138}$La decays have been enhanced and the environmental background has been further reduced in the coincidence spectrum.
Setting gates in the Clover spectrum on 788.7 keV and 1435.8 keV of $^{138}$Ba decays, allows us to
pick up the $\beta$ spectrum and the X-ray spectrum in the LaBr$_3$(Ce) detector,
as shown in Fig.~\ref{fig4}(a).The coincidence $\beta$ spectrum has triggered a precise study of $^{138}$La decay~\cite{Quarati2012Study,Giaz2015Measurement,Quarati2016Experiments,Quarati2016Reprint,Sandler2019Direct}, which is a second forbidden unique decay.
The $\beta$ and X-ray distributions are shown for comparison in the LaBr$_3$(Ce) spectrum.

%%%%%%%%%%%%%%%%%%%%%%%%%%%%%%%%%%%
\begin{center}
\includegraphics[width=8.0cm,clip=true,trim=0cm 0cm 0cm 0cm]{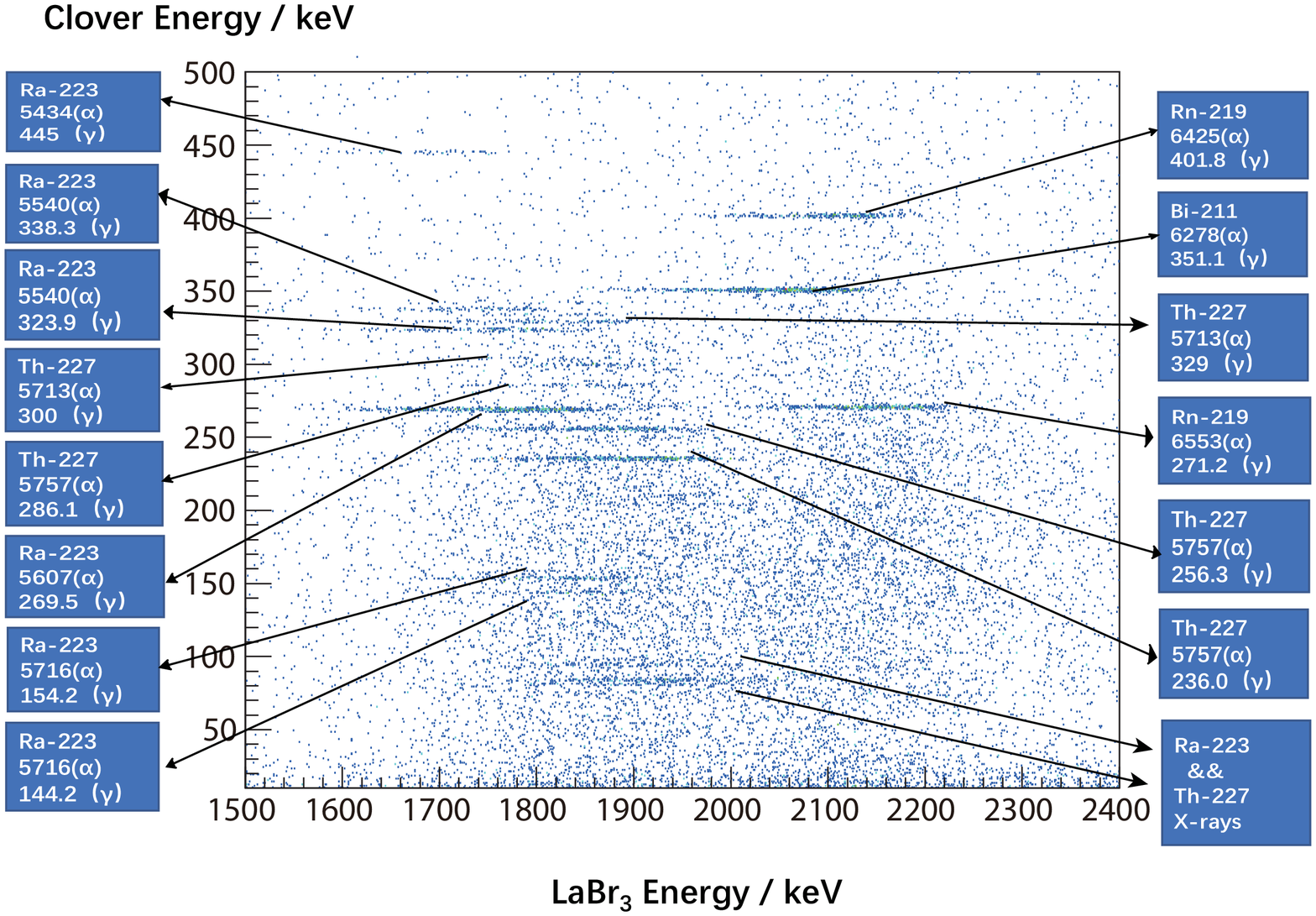}
\figcaption{\label{fig5}(Color online) Matrix of the LaBr$_3$(Ce) vs Clover. Listed are also the events of $\alpha-\gamma$ correlations and the identified radioactive isotopes. Both detectors are calibrated with characteristic $\gamma$-rays.
The events between 1.5 and 2.4 MeV of LaBr$_3$3(Ce) correspond to 5-7.4 MeV $\alpha$ particles of the $^{227}$Ac decay chain. See text for details.}
\end{center}

%%%%%%%%%%%%%%%%%%%%%%%%%%%%%%%%%%

Part of the coincidence events in the Clover and LaBr$_3$(Ce) detectors is displayed in Fig.~\ref{fig5}, while the relevant projection to the Clover in Fig.~\ref{fig6}.
The horizontal bands in Fig.~\ref{fig5} are traced back to the $\alpha$-$\gamma$ cascades.
The correlated $\gamma$ ray energies and their relative intensities are key to identify the $\alpha$ emitters.

The $\alpha$ emitters are identified to be $^{227}$Th, $^{223}$Ra, $^{219}$Rn, and $^{211}$Bi. This is consistent to previous investigations on the LaBr$_3$(Ce) with a size $\phi1'' \times1''$~\cite{Nicolini2007Investigation} and  LaCl$_3$(Ce) with a size $\phi25$mm $\times 25$mm)~\cite{Milbrath2005Characterization}. %
The $\alpha$ particles in Fig.~\ref{fig5} can be classed into 2 groups by sorting the energies, i.e., at  5.5-5.7 MeV (5.540, 5.434, 5.607 and 5.716 MeV $\alpha$ from $^{223}$Ra, 5.713 MeV $\alpha$ from $^{227}$Th), and 6.2-6.4 MeV (6.278 MeV $\alpha$ from $^{211}$Bi, and 6.425 MeV $\alpha$ from $^{219}$Rn).
The third peak in the single spectrum of LaBr$_3$(Ce) in Fig.~\ref{fig4}(a) is the 7.386 MeV $\alpha$ line
from the ground state of $^{215}$Po to the ground state of $^{211}$Pb, in which there is no cascade $\gamma$-ray.

%%%%%%%%%%%%%%%%%%%%%%%%%%%%%%%%%%%
\begin{center}
\includegraphics[width=9.0cm,clip=true,trim=0cm 0cm 0cm 0cm]{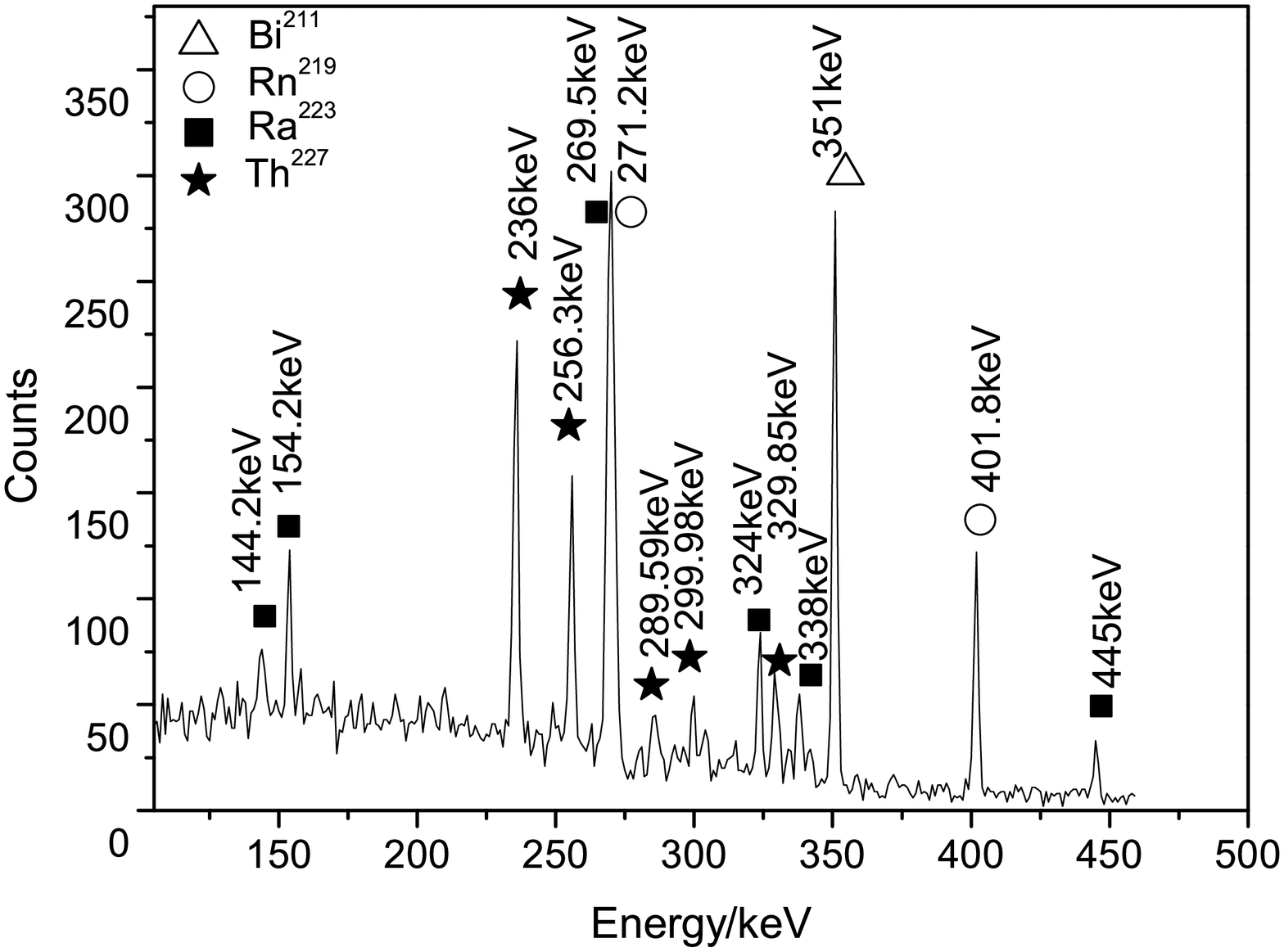}
\figcaption{\label{fig6}Projected $\gamma$-ray energy spectrum of Fig.~\ref{fig5} in the Clover. Labeled are the identified nuclei. }
\end{center}
%%%%%%%%%%%%%%%%%%%%%%%%%%%%%%%%%%

\section{Discussion}\label{dis}

In this section, we will deduce the activity of various radioactive contaminations embedded in the LaBr$_3$(Ce).
Since the radioactive contaminators are evenly distributed in the crystal, it is practically impossible to make a direct determination
because of a lack of an accurate efficiency calibration to both detectors.

Instead, in this work we develop a Monte Carlo model based on the GEANT4 version 10.4~\cite{S2002GEANT4,Allison2006Geant4,Allison2016Recent} tookit.
The setup includes the Clover, LaBr$_3$(Ce) detectors as well as its bracket, and all components of shielding system as shown in Fig.~\ref{fig3}.
$^{138}$La isotopes are evenly distributed in the  $\phi3^{''}\times3^{''} $ cylindrical LaBr$_3$ crystal. The density of LaBr$_3$ (Ce) crystal is set to be 5.08 g/cm$^3$~\cite{saintgobain}.
We employed the physics constructor class of G4EmStandardPhysics~\cite{Allison2016Recent}.
It includes various processes like the deposition of $\beta$ and $\gamma$ rays in the sensitive volumes of the detectors,
the occurrence of Compton scattering in the shield, and the characteristic X-rays induced from
the shield material.
The shielding materials are summarized in Table~\ref{tab:1}.
The activities of $^{138}$La and $^{227}$Ac decay chain contaminators are determined by reproducing the experimental spectra in both  the LaBr$_3$(Ce) and Clover detectors.

\begin{table*}[htbp]
%\caption{Informations of experimental targets.}
\centering\caption{Details of shielding materials of the LBS. Listed are the material layer, its inner radius, outer radius, and materials defined in GEANT4. }
%\label{tab:1}
%\vspace{0.2cm}
\label{tab:1}
\begin{tabular}{ccccc}
\hline\hline
Layer &  Inner radius/cm & Outer radius/cm &Material \\
\hline \rule{0em}{10pt}
 Iron        &     30      & 32     &    G4\_Fe               \\
 Lead        &     21.65   & 30     &    G4\_Pb              \\
 Copper      &     21.45   & 21.65  &    G4\_Cu                       \\
 Plexiglass  &     20.95   & 21.45  &    G4\_PLEXIGLASS                       \\

\hline \hline

\end{tabular}
\end{table*}

%%%%%%%%%%%%%%%%%%%%%%%%%%%%%%%%%%%%%%%%%%%

\subsection{Simulation of the self-counting LaBr$_3$(Ce)  }%Reproduce the 
%%%%%%%%%%%%%%%%%%%%%%%%%%%%%%%%%%%
\begin{center}
\includegraphics[width=8.0cm,clip=true,trim=0cm 0cm 0cm 0cm]{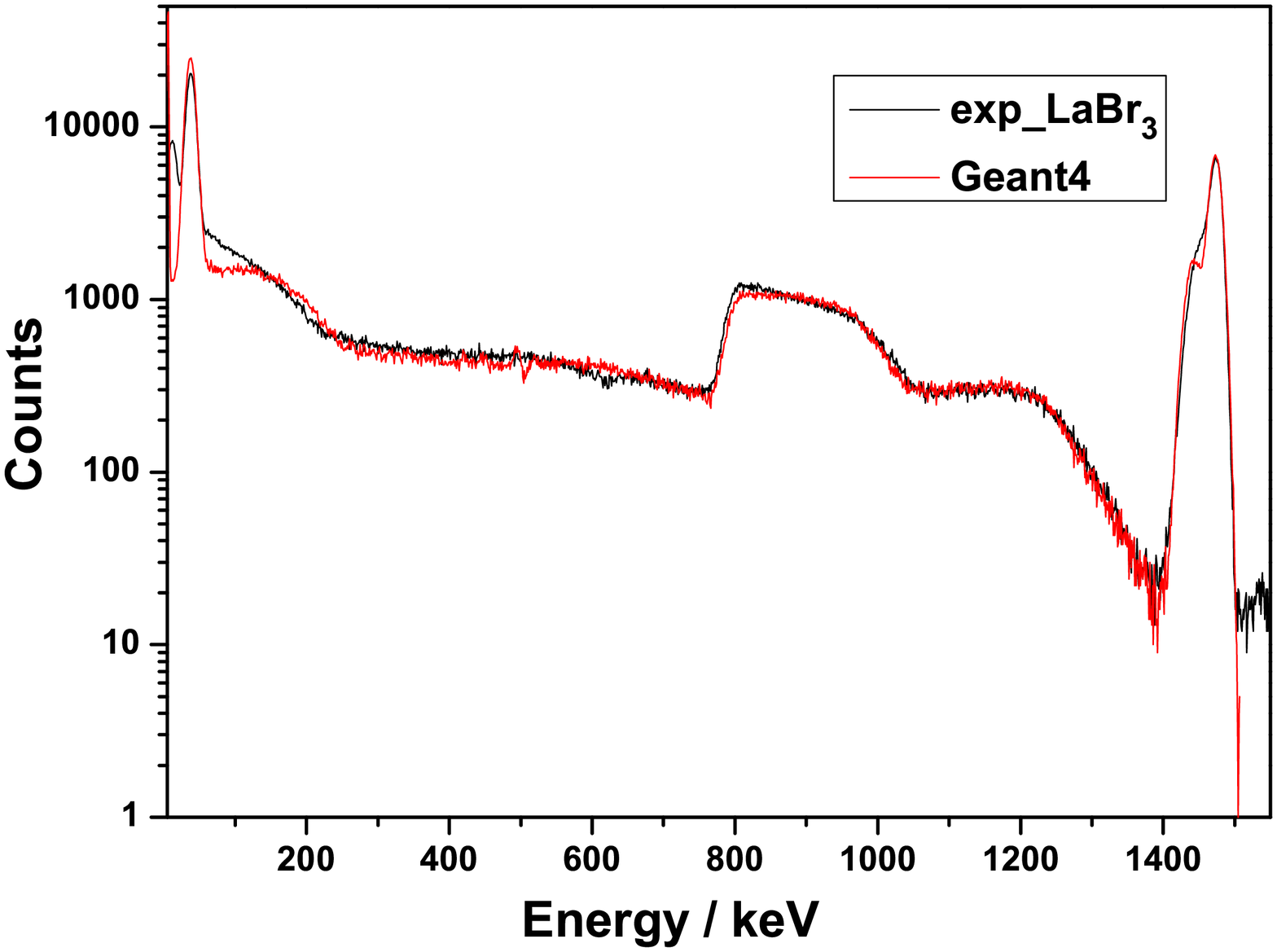}
\figcaption{\label{fig7}(Color online)Comparison of experimental (black line) and simulated (red line) self-counting energy spectrum in the LaBr$_3$  detector. The sampling of simulation has been scaled to the experimental data collected in 3200 seconds. }
\end{center}
%%%%%%%%%%%%%%%%%%%%%%%%%%%%%%%%%%
In our experiment, we set up the time stamp for each event in our DAQ system. This allows us to compare directly the simulation with the experimental data that are acquired in same measurement time, i.e., 3200 s.
The best fit to the experiment spectrum is found using the least square method when the activity of $^{138}$La 482(19) Bq, corresponding to 1.386 (55) Bq/cm$^{3}$. This corresponds to 1,543,500 decays of $^{138}$La in 3200 seconds in total.
 The uncertainty quoted here is due to the experimental statistics, detection efficiency and branching ratio of the $\gamma$-rays.
The experimental and simulated spectra are plotted in Fig.~\ref{fig7}. Nice agreement is seen except the low energy part up to about 150 keV.

It should be noted that in the above simulation we did not consider the contribution from $^{227}$Ac and its daughter nuclei.
This may partially account for the difference between the simulation and experiment spectra.
Another possible reason for the low energy deficiency in the simulation could be related to the insufficient understanding of $\beta$ decay of $^{138}$La. This has been discussed in Ref.~~\cite{Giaz2015Measurement,Quarati2016Experiments,Quarati2016Reprint,Sandler2019Direct}.

\subsection{Activities determined by using the Clover data}%Simulate the extraction efficiency of body source to extract activity of $^{138}$La }
An alternative way to deduce the $^{138}$La activity is to use the coincidence $\gamma$ rays information at 788.7 keV and 1435.8 keV in the Clover.
This would require an accurate efficiency calibration to the Clover detector using a volume source of the same volume as the LaBr$_3$(Ce) detector.
This is practically impossible.

In reality, we made a two-step optimization for the calibration~\cite{He2018Summing}.
In step one, we follow the standard routine for an efficiency calibration using standard radioactive point sources, $^{137}$Cs, $^{241}$Am, $^{54}$Mn, $^{88}$Y, $^{109}$Cd, $^{65}$Zn and $^{152}$Eu. These sources are selected to avoid possible true summing effect.
The point source was placed 3 cm from the front surface of the Clover detector.
We used the EFFIT program in the Radware package~\cite{Radware} to describe the
efficiency curve at the low-energy and high-energy regions separately.
%%%%%%%%%%%%%%%%%%%%%%%%%%%%%%%%%%%
\begin{center}
\includegraphics[width=8.0cm,clip=true,trim=0cm 0cm 0cm 0cm]{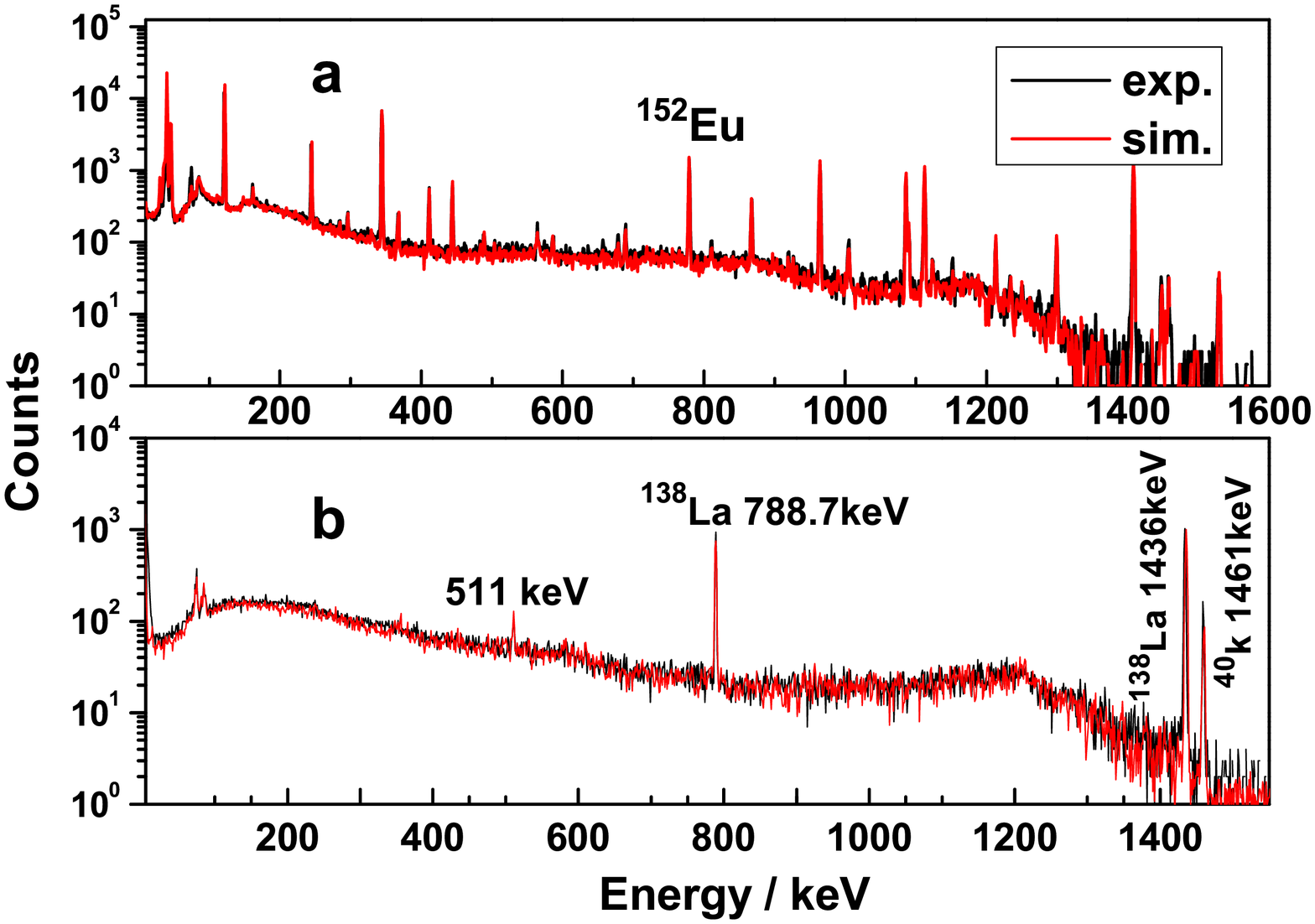}
\figcaption{\label{fig8}(Color online) Comparison of Clover spectrum (black line) with the simulation (red line) for the
$^{152}$Eu source (a) and the LaBr$_3$(Ce) crystal  (b). The sampling of simulation for LaBr$_3$(Ce) has been scaled to the experimental data collected in 3200 seconds  }
\end{center}

%%%%%%%%%%%%%%%%%%%%%%%%%%%%%%%%%%the low-background shielding system ,
%%%%%%%%%%%%%%%%%%%%%%%%%%%%%%%%%%%
\begin{center}
\includegraphics[width=8.0cm,clip=true,trim=0cm 0cm 0cm 0cm]{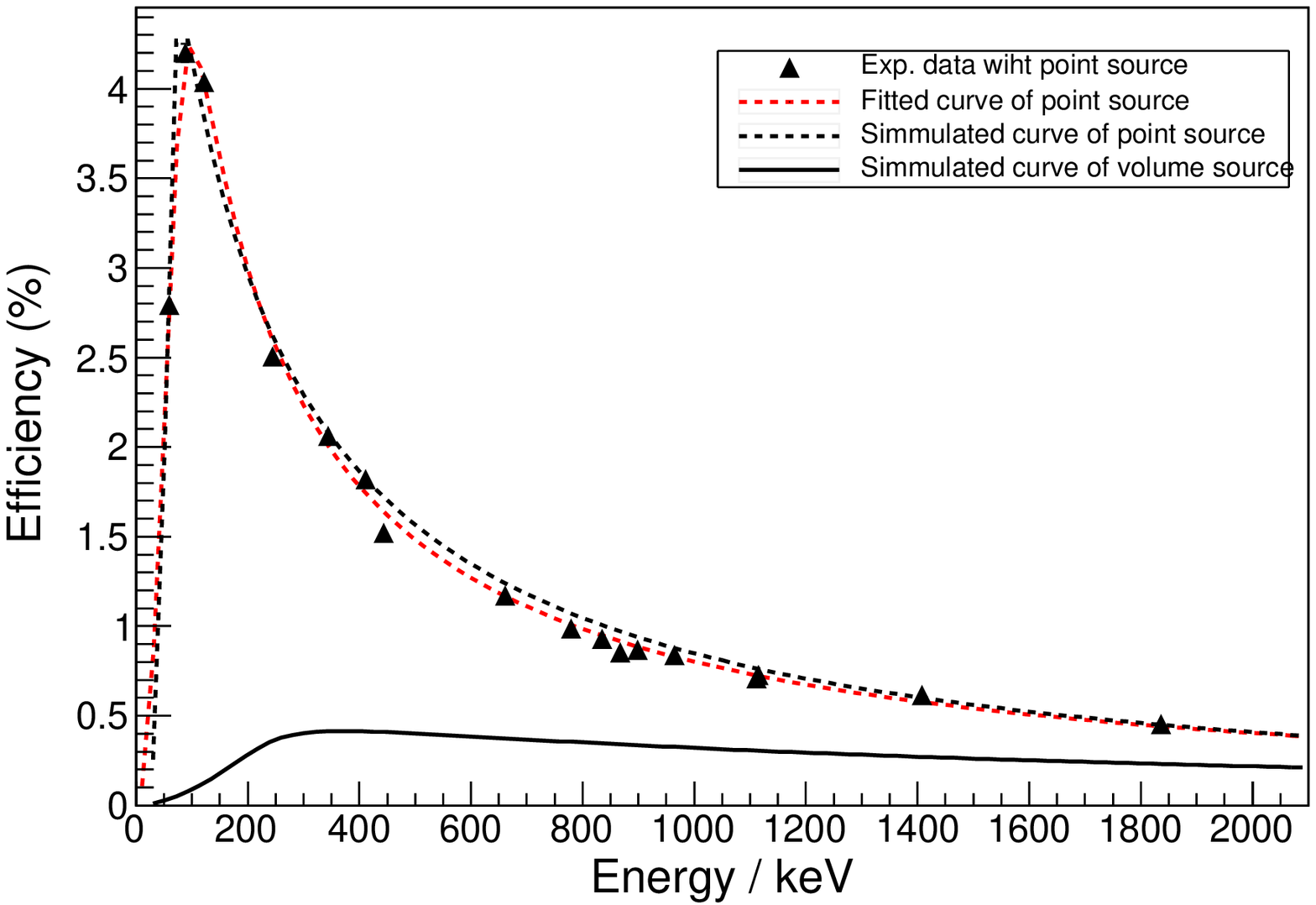}
\figcaption{\label{fig9}(Color online) The $\gamma$-ray detection efficiencies of Clover. The red dash line is the fitting curve to the experimental data (solid triangle) of standard point source, while the black dash line is the simulation efficiency under same detection configuration. The black solid line represents the detection efficiency when the source evenly distributed in the LaBr$_3$(Ce) crystal. }
\end{center}
%%%%%%%%%%%%%%%%%%%%%%%%%%%%%%%%%%

In this step, we optimized the thickness of the dead layer encapsulating the crystal by the least square method to reproduce best the Clover spectrum of $^{152}$Eu.
The best simulation results of $^{152}$Eu spectrum together with the experimental data are presented in Fig.~\ref{fig8}(a). The simulated detection efficiency curve is compared with the experimental results using standard point source, as shown in Fig.~\ref{fig9}.

In step two, we modeled the detection efficiency of the Clover with sources evenly distributed in the LaBr$_3$(Ce) crystal instead of point sources.
The simulated efficiency curve is shown in Fig.~\ref{fig9}. The efficiency of 788.7 keV and 1435.8 keV $\gamma$ rays are determined to be 0.00347(11) and 0.00257(10), respectively. In the simulation, the interaction of characteristic $\gamma$ rays with both the LaBr$_3$(Ce) and Clover detectors and even the shielding material have been taken into account.
\begin{table*}[htbp]
%\caption{Informations of experimental targets.}
\centering\caption{Activities of $^{211}$Bi, $^{219}$Rn, $^{223}$Ra.$^{227}$Th and $^{138}$La. Listed are the isotopes, the most distinct $\gamma$ rays, the deduced activities. Results from Ref.~\cite{Milbrath2005Characterization,Quarati2012Study}  are shown whenever available. }
\label{tab:2}
%\vspace{0.2cm}
%\label{systematic}
%\begin{threeparttable}
\begin{tabular}{ccccc}
\hline\hline

 Isotope &  $\gamma$-energy (keV)\tnote{1} & Absolute efficiency($\%$) &Activity (Bq/cm$^{3}$) &  Reference value (Bq/cm$^{3}$)\\
  \hline \rule{0em}{10pt}
  $^{211}$Bi  &     351.1   & 0.4347  &     0.0136(15)   & 0.032$^{ a}$      \\
  $^{219}$Rn  &     271.2   & 0.4117  &     0.0134(17)      &  /              \\
  $^{219}$Rn  &     401.8   & 0.4317  &     0.0116(16)   & 0.032$^{a}$              \\
  $^{223}$Ra  &     269.5   & 0.4130  &    0.0126(13)      & 0.025$^{a}$             \\
  $^{223}$Ra  &     445.0   & 0.4283  &    0.0127(15)   &   /                  \\
  $^{227}$Th  &     236.0   & 0.3840  &     0.0158(22) & 0.037$^{a}$                      \\
  \multicolumn{3}{c}{Sum activity of $\alpha$ contaminator} & 0.0545(35) & 0.126$^{a}$, 0.0443$^{b}$  \\
  \hline \rule{0em}{10pt}
  $^{138}$La  &     788.7   & 0.3467  &     1.451(58)       & /                \\
  $^{138}$La  &     1435.8   & 0.2567  &     1.437(63)      &  /                \\
  $^{138}$La  &\multicolumn{2}{c}{self-counting method} &     1.386(55)      &  /     \\
  \multicolumn{3}{c}{Avergage activity of $^{138}$La } &     1.425(59)      &   1.488$^{b}$     \\
  \hline \rule{0em}{10pt}&
  \multicolumn{2}{c}{Total activity} &     1.480(69)      &   1.532$^{b}$\\

\hline \hline

\end{tabular}
\begin{tablenotes}
        \footnotesize
        \item[a] $^{a}$Date from Ref.~\cite{Milbrath2005Characterization}
        \item[b] $^{b}$Date from Ref. ~\cite{Quarati2012Study}
      \end{tablenotes}
%\footnotesize{$^a$ The smallest spatial unit is county}\\
\end{table*}
As a result, we found that the Clover spectrum collected in 3200s is best reproduced when the total count of the 788.7 keV $\gamma$ ray is 1932 (46). The simulation together with the experimental data is shown in Fig.~\ref{fig8}(b).
%the counts of 789 keV $\gamma$ rays detected by Clover was 93033 (318) to reproduce the Clover spectrum,
The activity of $^{138}$La is thus extracted to be 505 (20) Bq, corresponding to 1.451 (58) Bq/cm$^3$.
The uncertainty takes into account the contributions from statistic, detection efficiency and branching ratio of the $\gamma$ rays in the decay of $^{138}$La. In the same way, the activity of $^{138}$La has been also deduced from the 1435.8 keV $\gamma$-ray, to be 500 (22) Bq, corresponding to 1.437 (63) Bq/cm$^3$.
Both are consistent with that determined from the self-counting of LaBr$_3$(Ce) detector, as shown in Table~\ref{tab:2}.

The above procedure can also be applied to evaluate the activities of the $^{227}$Ac decay chain contaminators.
In Fig.~\ref{fig6}, we identified the characteristic $\gamma$ rays at 351.1 keV ($^{221}$Bi), 271.2 keV and 401.8 keV ($^{219}$Rn),
269.5 keV and 445.0 keV ($^{223}$Ra), 236.0 keV ($^{227}$Th).
The deduced activities for $^{219}$Rn, $^{223}$Ra, $^{221}$Bi and $^{227}$Th are summarized in Table~\ref{tab:2}.
The sum activity of $\alpha$ contaminators is 0.0545(35) Bq/cm$^{3}$, which is a factor of 25 smaller than that of $^{138}$La.
More specifically, this resulted in a counting rate (including environmental background) of 237 counts/s for the $\gamma$-rays energy between 20 and 500 keV, 181 counts/s between 500 and 1.5 MeV, 27 counts/s above 1.5 MeV in our LaBr$_3$(Ce) detector.
The self-irradiation affects its application to low production experiments, in particular for potential cases with a count  rate of less than 450 counts/s.

\iffalse Ref.~\cite{Milbrath2005Characterization} reported the activities of $^{227}$Th , $^{221}$Bi, $^{219}$Rn , $^{223}$Ra for a LaCl$_{3}$(Ce)  detector with a crystal size of $\phi$25mm$\times$25mm
using a similar approach as the present work.
The activities of $^{227}$Ac decay chain contaminators in the B380 is of typically 40\% of those in Ref.~\cite{Milbrath2005Characterization}.
It shows the actinium impurity has been significantly reduced in the last years.
\fi

A pioneering study~\cite{Milbrath2005Characterization} of the LaCl$_{3}$(Ce)
detector with a crystal size of  $\phi$25 mm $\times$25 mm shows
a contamination level of 1.3 $\times 10^{-13}$ $^{227}$Ac decay chain atoms/La atom.
The identified total $\alpha$ activity ($^{227}$Th , $^{221}$Bi, $^{219}$Rn and $^{223}$Ra)
is 0.126  Bq/cm$^{3}$ and each contribution is summarized in Table ~\ref{tab:2}.
A recent study~\cite{Quarati2012Study} further investigated the internal radiation of LaBr$_3$(Ce) detectors with various sizes of crystals. Average activities of $^{138}$La and $^{227}$Ac decay chain are determined to be 1.488 and 0.044 Bq/cm$^{3}$, respectively.
In the present study of the B380 type, the $^{227}$Ac decay chain atoms/La atom amounts to be 5.8$\times 10^{-15}$. This is almost two orders of magnitude smaller than that in Ref.~\cite{Milbrath2005Characterization}.
The activities of $^{227}$Ac decay chain contaminators are of typically 40\% of that in Ref.~\cite{Milbrath2005Characterization}
and are similar to that in Ref.~~\cite{Quarati2012Study}. %. \textbf{MAKE SURE \cite{Milbrath2006Contamination,Iltis2006Lanthanum}.}
This indicates that the actinium impurity has been significantly reduced in the last decade.

\section{Summary}\label{sum}

In this work, we performanced a coincidence measurement using a Clover detector, and identified the internal radioactive nuclei of the B380 LaBr$_3$(Ce) detector.
We determined the activity of $^{138}$La, $^{211}$Bi, $^{219}$Rn, $^{223}$Ra and $^{227}$Th by combining the coincidence spectra with Geant4 simulations.\\
\indent The sum activities of  $^{138}$La and $^{138}$Ac decay chain are around 1.425 (59) and 0.0545 (35) Bq/cm$^3$, respectively.
These data are useful in designing detector setups based on the LaBr$_3$(Ce), in particular for the purpose of low count rate experiments.

\end{multicols}
\vspace{-1mm}
\centerline{\rule{80mm}{0.1pt}}
\vspace{2mm}
\begin{multicols}{2}

\end{multicols}

\clearpage
\end{CJK*}
\end{document}